\documentclass[twocolumn,showpacs,preprintnumbers,amsmath,amssymb, prb, superscriptaddress]{revtex4}

\usepackage{graphicx}% Include figure files
\usepackage{dcolumn}% Align table columns on decimal point
\usepackage{bm}% bold math

\usepackage{color,ulem}

\begin{document}
\title{Angle-dependence of the frequency correlation in random photonic media: the diffusive regime and its breakdown near localization.}
\author{O. L. Muskens}
\affiliation{SEPnet and Department of
Physics and Astronomy, University of Southampton, Highfield SO17
1BJ, Southampton, UK.}\email{O.Muskens@soton.ac.uk}
\affiliation{Center for Nanophotonics, FOM Institute for Atomic
and Molecular Physics AMOLF, Kruislaan 407, 1098 SJ Amsterdam, The
Netherlands.}
\author{T. van der Beek}
\affiliation{Center for Nanophotonics, FOM Institute for Atomic
and Molecular Physics AMOLF, Kruislaan 407, 1098 SJ Amsterdam, The
Netherlands.}
\author{A. Lagendijk}
\affiliation{Center for Nanophotonics, FOM Institute for Atomic
and Molecular Physics AMOLF, Kruislaan 407, 1098 SJ Amsterdam, The
Netherlands.}

\date{\today}

\begin{abstract}
The frequency correlations of light in complex photonic media are of
interest as a tool for characterizing the dynamical aspects of light diffusion. We demonstrate here that the frequency correlation shows a pronounced angle dependence both in transmission and in
reflection geometries. Using a broadband white light
supercontinuum, this angle dependence is characterized and
explained theoretically by a combination of propagation effects
outside the medium and coherent backscattering. We report a strong
dependence of the coherent backscattering contribution on the
scattering strength which cannot be explained by the diffusion theory. Our results indicate that coherent backscattering of the frequency correlation forms a sensitive probe for the breakdown of the diffusive regime near localization.
\end{abstract}
 %}%} \narrowtext
\maketitle

\section{Introduction}

Nanostructured photonic media, such as photonic crystals, metamaterials, and random media, are of interest their potential of achieving nanoscale control over the flow and emission of light. The wave nature of light results in a range of interference phenomena, which modify the propagation of light through complex photonic materials, in analogy with the theory of electrons in solids. Interference effects lead to new correlations between optical fields at different angles and frequencies after interaction with the medium.\cite{Feng94} For random or partially disordered photonic media, methods have appeared that allow for a statistical interpretation of the spectral width of the transmission and reflection correlations in terms of the distribution of dwell times of light inside the medium.\cite{Genack87,Genack90,Feng94,Albada90,DeBoer92} The short-range frequency correlation, known as $C^{(1)}$, which is present in the spectrum after interaction with the medium, has evolved into a broadband tool for characterizing the dynamic transport parameters of photonic nanomaterials.\cite{muskensoptlett09}

Here, we explore the angle-dependence of the frequency correlation both in transmission and reflection geometries. Previous theoretical and experimental studies of the $C^{(1)}$ frequency correlation have not considered its dependence on scattering angle.\cite{DeBoer92} The angle-dependence of the angular correlation function in transmission was studied by Li and Genack \cite{Li94}, who showed that the far-field and near-field correlations form a Fourier-transform pair. We derive a similar relationship for the frequency correlation, which however is shown to produce an angle-dependent narrowing of the correlation width. In addition, we investigate for the first time interference contributions to the frequency correlation in backscattering resulting from coherent backscattering of light. For photonic media, time-reversal symmetry dictates that for every path contributing to the reflectance, a time-reversed path is present that interferes constructively in exact backscattering. This effect, which is a manifestation of weak localization, is known in photonics as the coherent backscattering (CBS) effect.\cite{Albada85,WolfPRL85} The constructive interference in coherent backscattering results in a cone of enhanced intensity versus angle around the backscattering direction, which is of order unity in intensity for all scattering strengths. The angular extent of the CBS cone increases proportional to the photonic strength $(k_{e} \ell_B)^{-1}$ of the medium, where $k_{e}$ denotes the wavevector of light in the medium and $\ell_B$ denotes the (Boltzmann) transport mean free path.\cite{AkkermansBook}

Effects of coherent backscattering have been observed in the time-correlation function of colloidal suspensions\cite{Qu88} and in the angular correlation function ('memory effect').\cite{Berkovits90,Feng94} However, the direct contribution of time-reversed light paths has never been investigated for the frequency correlation. A pronounced angle-dependence of this correlation can be expected from the relationship between the time duration of a short pulse after diffuse propagation and the spectral correlation function\cite{Genack90}, combined with the observation of angle-dependent changes in the pulse propagation time in scattering media.\cite{Vreeker89} We present here both a theoretical and experimental description of this effect, using a series of strongly scattering scattering media with scattering strengths ranging from the diffuse scattering to the strongly photonic regime. By varying the scattering strength, we study the breakdown of the diffusion approximation for the correlations in coherent backscattering. We report results showing that the frequency correlation depends on scattering strength in a new and previously unexplored way, and which indicate that the frequency correlation may provide a more sensitive means for accessing long light paths compared to the conventional intensity CBS effect.\cite{SchuurmansPRL99} Generally, the frequency correlation can be used to obtain information on the distribution of photon dwell times complementary to time-resolved experiments, which are usually not easily extended to angle-dependent measurements.\cite{Toninelli08,StoerzerPRL06,JohnsonPRB03,Vellekoop} In addition, we discuss the possibility that the combination of CBS and frequency correlations yields additional higher-order correlation effects, which are unique for the four-field correlation and are not found in conventional intensity measurements.

\section{Theory}
\subsection{Angle-dependent frequency correlations in transmission}
\label{sec:theoryT}
Following the general definition\cite{Feng94}, the short-range frequency correlation function
$C^{(1)}_{\mathbf{\hat u},\mathbf{\hat a}}(\omega,\Omega)$ between a single scattering direction $\mathbf{\hat u}$ and
a single incident direction $\mathbf{\hat a}$ is given by
\begin{equation}\label{eq:C1general}
C^{(1)}_{\mathbf{\hat u},\mathbf{\hat
a}}(\omega,\Omega)=\frac{\left|\left< t_{\mathbf{\hat
u},\mathbf{\hat a}}(\omega+\Omega) t^*_{\mathbf{\hat
u},\mathbf{\hat a}}(\omega)\right>\right|^2}{\left<
|t_{\mathbf{\hat u},\mathbf{\hat
a}}(\omega+\Omega)|^2\right>\left< |t_{\mathbf{\hat
u},\mathbf{\hat a}}(\omega)|^2\right>} \,
\end{equation}
with $t_{\mathbf{\hat u},\mathbf{\hat a}}$ the transmission coefficient for the electric field
and $\Omega$ the frequency difference between two fields. The
short-range frequency correlation has been well studied both
experimentally and
theoretically.\cite{Genack90,Albada90,DeBoer92,Feng94} We follow
here the derivation by De Boer et al. in a slightly
different notation, where we make use of the 'intensity on a
scatterer' $\mathcal{I}$, i.e. the intensity at the last scatterer
in the medium.\cite{DeBoer92,deBoerPhD} The intensity $\mathcal{I}$ is related to the source $S$ by an
intensity propagator $H$ which contains the 4-point ladder vertex
with attached incoming Green's functions. After Fourier
transformation of the tranverse coordinate, $\mathcal{I}$ can be
written as
\begin{eqnarray}\label{eq:Ifirst}
\mathcal{I}(\mathbf{q}_\perp,z;t) &=& \frac{16 \pi^2}{\ell_B}\int
{\rm
d}z'\int {\rm d}t' \\
&&\times H(\mathbf{q}_\perp,z,z';t-t') S(\mathbf{q}_\perp, z';t').
\nonumber
\end{eqnarray}

Equation~\ref{eq:Ifirst} can be readily extended to combinations
of fields with different frequencies by modifying the source $S$.
In the following, we consider $\mathcal{\tilde  I}$ as a complex
field-field correlator instead of a real-valued intensity. The
incident light beam is described by a Gaussian intensity profile
with waist $\rho_0$, which is injected into the scattering medium
at an injection depth $z_i$. For the frequency correlation, we
consider a mixed-frequency source containing two oscillating
fields with frequencies $\omega$ and $\omega+\Omega$. This yields
an oscillating phase $e^{i\Omega t}$ in the time response. After
Fourier transformation, the injected source field profile, $\tilde
S$, is given by
\begin{equation}
\label{eq:Source} \tilde S(\mathbf{q}_{\perp}, z;t)= \pi \rho_0^2 I_0 \ell_B
\delta(z-z_i) e^{-\frac{1}{4}|\mathbf{q}_\perp|^2
\rho_0^2} e^{i\Omega t}\, .
\end{equation}

\begin{figure}[t]
\includegraphics[width=4cm]{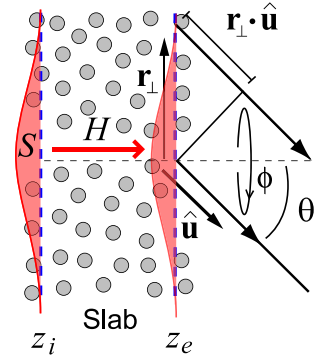}
\caption{\label{fig:scheme} (color online) Schematic picture of light transport
through a diffusive slab, indicating injection and extraction
planes at $z_i$ and $z_e^T$, Gaussian source profile (shaded area,
red), and the additional path length difference $\xi$ outside the
medium.}
\end{figure}

The propagator $\tilde H$ of a finite slab is considered, which will be
most convenient for further use when written as a function of
frequency
\begin{equation}\label{eq:Hfourier}
\tilde H(\mathbf{q}_\perp, z,z';t)=\frac{1}{2\pi}\int {\rm d}\Omega
e^{i\Omega t} \tilde H(\mathbf{q}_\perp, z,z';\Omega) \,
\end{equation}
with\cite{DeBoer92}
\begin{eqnarray}
\label{eq:Hslab} \tilde H(\mathbf{q}_\perp, z,z';\Omega)&=&\frac{v_E}{4
\pi
\ell_B D} \\
&& \times
\frac{\sinh[(z'+\tau_0) \tilde\eta]\sinh[(L-z-\tau_0)\tilde\eta]}{\tilde \eta\sinh[L\tilde\eta]}
\nonumber \, ,
\end{eqnarray}
where $\tilde \eta=(|\mathbf{q}_\perp|^2 + 1/D\tau_a - i\Omega/D)^{1/2}$,
$D$ being the diffusion constant of the light. The effect of internal reflections can be
included in $\tilde H$ following for example Zhu et al.\cite{Zhu91} The
diffuse propagator $\tilde H$ in principle depends on the optical
frequency $\omega$ through the dispersion of the parameters $v_E$,
$\ell_B$, $\tau_a$, and $D$ in Eq.~(\ref{eq:Hslab}). However, in
our experiments we will consider these parameters constant as
their variation with frequency in the visible spectrum has been
shown to be smooth for the materials under study
\cite{JohnsonPRB03,muskensOE08, muskensoptlett09,muskensNanoLett09}.

The expression for $\mathcal{\tilde I}(\mathbf{q}_\perp,z,t)$ follows
from Eq.~(\ref{eq:Ifirst})$ - $(\ref{eq:Hfourier}) as
\begin{eqnarray}
\mathcal{\tilde I}(\mathbf{q}_\perp,z;t)&=& 16 \pi^3 \rho_0^2 I_0  \int
{\rm d}t' \int {\rm d}\Omega e^{i\Omega (t-t')}
\tilde H(\mathbf{q}_\perp, z,z_i;\Omega) \nonumber \\
&&\times e^{i\Omega t'} e^{-\frac{1}{4}|\mathbf{q}_\perp|^2 \rho_0^2} \nonumber \\
&=& \frac{1}{2\pi} \int {\rm d}\Omega e^{i \Omega
t}\mathcal{\tilde I}(\mathbf{q}_\perp,z;\Omega) \, ,
\end{eqnarray}
where the second line is obtained by the integration over $t'$ of
$e^{i (\Omega'-\Omega) t'}$, yielding $\delta(\Omega-\Omega')$.
Here the final expression for $\mathcal{\tilde
I}(|\mathbf{q}_\perp|,z;\Omega)$ is
\begin{eqnarray}
\label{eq:Iscatomega} \mathcal{\tilde I}(\mathbf{q}_\perp,z;\Omega)&=&\frac{8\pi^3 \rho_0^2 I_0 v_E}{\ell_B D}e^{-\frac{1}{4}|\mathbf{q}_\perp|^2 \rho_0^2} \\
&&\times \frac{\sinh[(z_i+\tau_0)\tilde \eta]\sinh[(L-z-\tau_0)\tilde \eta]}{\tilde \eta\sinh[L\tilde \eta]}\,
. \nonumber
\end{eqnarray}

The transmitted intensity in the far-field can be obtained by
attaching outgoing Green's functions to the mixed-field correlator
$\mathcal{\tilde I}$. In the most simple approximation, we assume
all the energy to be emitted from an ejection plane
$z_e^T=L-2\tau_0-\frac{2}{3}\ell_B$ with a weight of the ejection
function given by $\ell_B/16\pi$.\cite{deBoerPhD,AkkermansBook}
From the ejection plane, spherical waves are emitted into the
far-field. Since the detector collects intensity in an angular
range, we write down the expression of the far-field intensity in
direction $\mathbf{\hat u}$ at a wavefront emitted under an angle
$\theta$ with the normal. This is given by the integral of
intensities over all positions $\mathbf{r}_\perp$ on the ejection
plane times the Green functions containing the corresponding path
length difference
$\xi=\mathbf{r}_\perp\cdot\mathbf{\hat{u}}=|\mathbf{r}_\perp|\sin\theta
\cos\phi$, as shown in Fig.~\ref{fig:scheme}
\begin{equation}
\tilde g_{\mathbf{\hat{u}}}(r_0,\mathbf{r}_\perp;\omega) \approx
-\frac{e^{i \frac{\omega}{c_0} (r_0+\mathbf{r}_\perp \cdot
\mathbf{\hat{u}})}}{4 \pi r_0} \, ,
\end{equation}
where $r_0$ denotes the distance on the optical axis from the
detector to the center of the ejection plane (i.e.
$\mathbf{r}_\perp=0$) and $\phi$ is the azimuthal angle of
$\mathbf{r}_\perp$ with the radiation plane including $\theta$.

\begin{figure}[t]
\includegraphics[width=7.0cm]{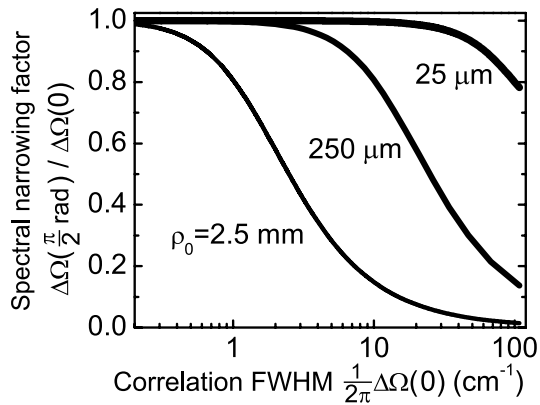}
\caption{\label{fig:narrowing} Calculated spectral narrowing of
the frequency correlation $\Delta \Omega(\frac{\pi}{2})/\Delta \Omega(0)$ against frequency correlation width $\Delta \Omega(0)/2\pi$, for various values of the incident beam
width $\rho_0$ of 25~$\mu$m (thick line), 250~$\mu$m (line) , and
2.5~mm (thin line).}
\end{figure}

The angle-dependent far-field intensity at a distance $r_0$ is
obtained by attaching the outgoing Green's functions to the
near-field intensity Eq.~(\ref{eq:Iscatomega}). This results in a
mixed-frequency bistatic coefficient\cite{Ishimaru} in transmission and for
perpendicular incidence, $\tilde \gamma^{T}_{\mathbf{\hat
u},\mathbf{\hat 0}}(\Omega)$, given by
\begin{widetext}
\begin{eqnarray}
\label{eq:Ifarangle} \tilde \gamma^T_{\mathbf{\hat u},\mathbf{\hat
0}}(\Omega)&=&\frac{4 \pi r_0^2}{I_0A}\frac{\ell_B}{16\pi} \int
{\rm d}\mathbf{r}_\perp \nonumber \tilde
g_{\mathbf{\hat{u}}}(r_0,\mathbf{r}_\perp;\omega)  \tilde
g^*_{\mathbf{\hat{u}}}(r_0,\mathbf{r}_\perp;\omega+\Omega)
\mathcal{\tilde I}(\mathbf{r}_\perp,z_e^T,
\Omega) \nonumber \\
&=&\frac{\ell_B e^{i \frac{\Omega}{c_0} r_0}}{64 \pi^2 I_0 A} \int
{\rm d} \mathbf{r}_\perp e^{i \frac{\Omega}{c_0} \mathbf{r}_\perp
\cdot \mathbf{\hat u}} \mathcal{\tilde
I}(\mathbf{r}_\perp,z_e^T;\Omega) \, ,
\end{eqnarray}
\end{widetext}
where $A$ denotes the detector area following the definition of
the bistatic coefficient.\cite{Ishimaru} The second line of
Eq.~(\ref{eq:Ifarangle}) can be identified as a Fourier transform
of $\mathcal{\tilde I}(\mathbf{r}_\perp,z_e^T,\Omega)$ to a
transverse momentum coordinate $\frac{-\Omega \sin
\theta}{c_0}\mathbf{\hat u}_\perp$, where $\mathbf{\hat u}_\perp
\sin \theta$ is the projection of the scatting vector
$\mathbf{\hat u}$ onto the ejection plane. This Fourier
relationship between the near-field and the far field spectral
correlation follows the van Cittert-Zernike theorem and is an
example of a more general theorem dealing with spectral invariance
of light on propagation \cite{Wolf86}. A similar Fourier
relationship holds for the angular correlation function, as was
shown by Li and Genack.\cite{Li94}. The Fourier transform results
in a final expression for the scattering bistatic coefficient
\begin{equation}
\label{eq:Ifarangle2} \tilde \gamma^T_{\mathbf{\hat
u},\mathbf{\hat 0}}(\Omega)=\frac{\ell_B e^{i \frac{\Omega}{c_0}
r_0}}{64 \pi^2 I_0 A} \mathcal{\tilde I}(-\frac{\Omega \sin \theta
}{c_0}\mathbf{\hat u}_\perp ,z_e^T;\Omega) \, ,
\end{equation}
with $\mathcal{\tilde I}$ the near-field correlator in momentum
space, Eq.~(\ref{eq:Iscatomega}). The correlation function of
Eq.~(\ref{eq:C1general}) can be calculated by identifying that
$\left< t_{\mathbf{\hat u},\mathbf{\hat 0}}(\omega+\Omega)
t^*_{\mathbf{\hat u},\mathbf{\hat 0}}(\omega)\right>\equiv \tilde
\gamma^{T}_{\mathbf{\hat u},\mathbf{\hat 0}}(\Omega)$, from which
the correlation in transmission follows as
\begin{eqnarray}\label{eq:c1t}
C^{(1)}_{\mathbf{\hat u},\mathbf{\hat 0}}(\Omega)&=& \frac{|\tilde
\gamma^T_{\mathbf{\hat u},\mathbf{\hat
0}}(\Omega)|^2}{\left|\tilde \gamma^T_{\mathbf{\hat
u},\mathbf{\hat 0}}(0)\right|^2 } \, .
\end{eqnarray}
Equation~(\ref{eq:Ifarangle2}) shows that the far-field frequency
correlation gains an additional angle-dependent dephasing which
can be written as a transverse momentum which is proportional to
the frequency difference $\Omega/c_0$ and the angle $\theta$.

The spectral correlation of Eq.~\ref{eq:c1t} has a characteristic spectral width $\Delta \Omega(\theta)$, which in the forward direction is approximately given by $\Delta \Omega(0)/2 \pi \simeq 1.46 D/L^2$.\cite{Genack90} For nonzero angles, the spectral width is reduced, as is shown in Fig.~\ref{fig:narrowing} where we have calculated the spectral narrowing $\Delta \Omega(\pi/2)/\Delta \Omega(0)$ as a function of the full-width at half maximum of the spectral correlation function. Figure~\ref{fig:narrowing} shows that the spectral narrowing depends strongly on the combination of the incident beam size $\rho_0$ and the initial spectral width. For typical beam sizes of around $25- 250$~$\mu$m, the narrowing effect is pronounced for correlation widths in the range $10-100$~cm$^{-1}$. This frequency width corresponds to dwell times $T_d \simeq \pi/\Delta \Omega$ in the range 0.1-1~ps, or several hundreds to thousands of scattering events. This regime applies to many types of photonic nanomaterials with thickness in the range of hundreds of nanometers to several micrometers, demonstrating that the angle dependence of the frequency correlation in Eq.~(\ref{eq:c1t}) is a sizeable effect in many situations.

\subsection{Angle-dependent frequency correlations in reflection}
\label{sec:theoryR}

While the $C^{(1)}$ frequency correlations in transmission are
governed by long light paths scaling with the slab thickness $L$,
the correlation in reflection is governed by paths of the order of
the mean free path $\ell_B$. In addition, the reflection geometry
is characterized by the presence of weak localization in the
backscattering direction.

\begin{figure}[t]
\includegraphics[width=8.5cm]{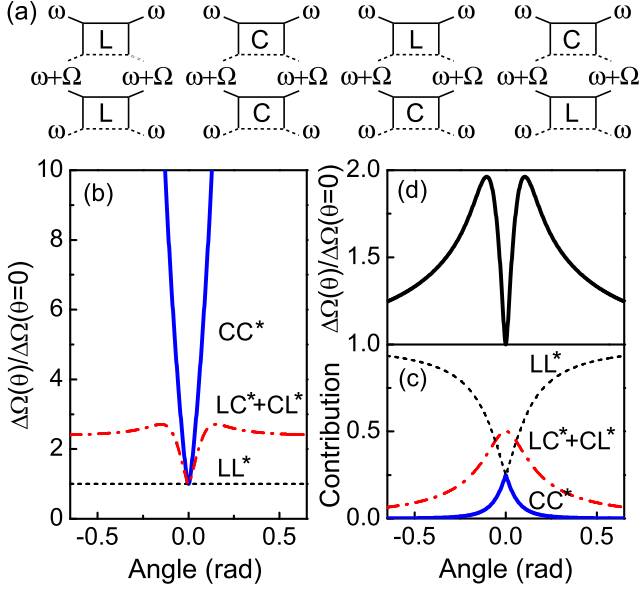}
\caption{\label{fig:c1cbscalc} (color online) (a) Four-field diagrams
contributing to the frequency correlation
$C^{(1)}_R(\omega,\Omega)$, from left to right LL*, CC*, LC*,
and CL*. (b) Frequency full-width at half-maximum $\Delta
\Omega$, normalized to width at backscattering $\theta=0$, for the
different diagram contributions. (c) Relative contribution of various diagrams to the total $C^{(1)}$ correlation. (d)
Total calculated broadening of correlation function against
backscattering angle.}
\end{figure}

We start again from the frequency correlation of
Eq.~(\ref{eq:C1general}), where this time we consider reflection
coefficients $r_{\mathbf{\hat
u},\mathbf{\hat 0}}$ for the electric field. For the case of
reflection, we expand the numerator of Eq.~(\ref{eq:C1general}),
analogous to Berkovits and Kaveh \cite{Berkovits90}, into
combinations of diffusive and maximally-crossed multiple
scattering contributions, respectively denoted here by L
(diffuson) and C (Cooperon) [cf.
Fig.~\ref{fig:c1cbscalc}(a)].\cite{AkkermansBook} The spectral
correlation function consists of four diagrams, consisting of two
diffusons (LL*), two Cooperons (CC*), and two mixed-diagram
contributions (LC* and CL*). We write the correlation as the sum
of two mixed-frequency bistatic coefficients
$\gamma^l_{\mathbf{\hat u},\mathbf{\hat 0}}(\omega,\Omega)$ and
$\gamma^c_{\mathbf{\hat u},\mathbf{\hat 0}}(\omega, \Omega)$
\begin{equation}
\label{eq:c1bistatic} C^{(1)}_{\mathbf{\hat u},\mathbf{\hat
0}}(\omega,\Omega) = N[|\gamma^l_{\mathbf{\hat u},\mathbf{\hat
0}}(\omega,\Omega)+\gamma^c_{\mathbf{\hat u},\mathbf{\hat
0}}(\omega, \Omega)|^2] \, ,
\end{equation}
where $N$ is a normalization factor given by
$N=\frac{1}{4}\gamma^l_{\mathbf{\hat u},\mathbf{\hat
a}}(\omega,0)^{-2}$. Analogous to the derivation in transmission,
the bistatic coefficient for the ladder contribution can be
written as
\begin{equation}\label{eq:gammal}
\tilde \gamma^{l}_{\mathbf{\hat u},\mathbf{\hat 0}}(\omega,\Omega) =
\frac{\ell_B e^{i \frac{\Omega}{c_0} r_0}}{64 \pi^2 I_0 A}
\mathcal{\tilde I}(-\frac{\Omega \sin \theta }{c_0}\mathbf{\hat
u}_\perp ,z_e^R;\Omega) \, ,
\end{equation}
where $z_e^R=\frac{2}{3}\ell$ denotes the ejection plane in
reflection, based on the average and the weight of the ejection function as defined in Ref.~\onlinecite{deBoerPhD}. For the case of the coherent backscattering contribution, the
solution is obtained by adding the incident and outgoing plane waves on positions $\mathbf{r}_\perp$ , $\mathbf{r}_\perp'$ \cite{AkkermansBook}
\begin{eqnarray}\label{eq:gammac}
\tilde \gamma_{c}(\mathbf{q}_\perp, \omega,\Omega)&&= \frac{\ell_B}{4 A
\mu_i} \int \int {\rm d}^2\mathbf{r}_\perp' {\rm
d}^2\mathbf{r}\,_\perp e^{-i \mathbf{q}_\perp \cdot
(\mathbf{r}_\perp-
\mathbf{r}_\perp')} \nonumber \\
&& \times \tilde H(\mathbf{r}_\perp-\mathbf{r}_\perp', z_i,
z_e, \Omega) \tilde S(\mathbf{r}_\perp',z_i;\Omega) \, ,
\end{eqnarray}
where $\mathbf{q}_\perp=k_0 \mathbf{\hat u}_\perp \sin \theta$ denotes the transverse wavevector difference of the incident and outgoing waves. Following the derivation of Sec.~\ref{sec:theoryT}, we find a similar expression for the bistatic coefficient with the addition of a wavevector $\mathbf{q}_\perp$
\begin{equation}\label{eq:gammac}
\tilde \gamma^{c}_{\mathbf{\hat u},\mathbf{\hat 0}}(\omega,\Omega) =
\frac{\ell_B e^{i \frac{\Omega}{c_0} r_0}}{64 \pi^2 I_0 A}
\mathcal{\tilde I}(\mathbf{q}_\perp-\frac{\Omega \sin \theta}{c_0}\mathbf{\hat
u}_\perp ,z_e^R;\Omega) \, .
\end{equation}

The angle-dependent frequency width of the various
diagrammatic contribution to the frequency correlation is shown in
Fig.~\ref{fig:c1cbscalc}(b), while their relative amplitude
contributions to the total correlation is shown in
Fig.~\ref{fig:c1cbscalc}(c). In exact backscattering ($\theta=0$),
$\gamma_c$ equals $\gamma_l$, and the contributions of the four
combinations LL*, CC*, CL*, and LC* are equal. For nonzero
wavevector $\mathbf{q}_\perp$, the CBS contributions broaden
spectrally due to the reduced weight of long light paths [cf.
Fig.~\ref{fig:c1cbscalc}(b)]. However, the CBS contribution
decreases rapidly in intensity for large scattering angles [cf.
Fig.~\ref{fig:c1cbscalc}(c)], resulting in recovery of the purely
diffusive correlation at large angles. The resulting frequency
broadening of the total correlation, including the contributions
of all diagrams, is shown in Fig.~\ref{fig:c1cbscalc}(d).

The angle dependence in the coherent backscattering frequency
correlation shows similarities to the time-correlation function of weakly scattering colloidal
suspensions in backscattering.\cite{Qu88,Ishii97} These effects are related because of the presence of the same path length
distribution in the dephasing caused by moving particles.
However, whereas the time-correlation is governed by the slow movement of scattering colloids, the frequency correlation in our case
directly relates to the photon transport time in the random medium with quenched disorder.

\section{Methods}

\begin{figure}[t]
\includegraphics[width=7.5cm]{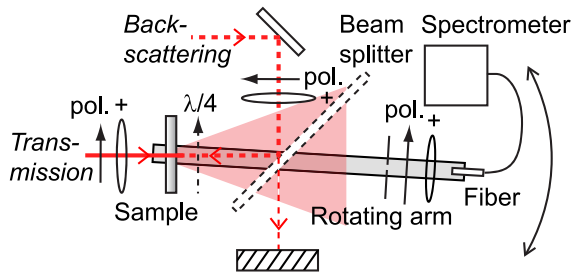}
\caption{\label{fig:setup} (color online) Experimental setup for white-light, angle-dependent frequency correlations in transmission and backscattering geometries. Dashed lines indicate components which are only present in backscattering configuration.}
\end{figure}

Correlations were measured in the frequency domain using a broadband technique developed by us as described in Ref~\onlinecite{muskensoptlett09}. Spectra containing individual frequency speckles were collected as a function of scattering angle using a rotation stage and a fiber-coupled grating spectrometer, as shown in Fig.~\ref{fig:setup}. Polarization filters were used to select linear polarization channels for the incident and detected light. In the coherent backscattering experiments, a quarter wavelength plate was placed directly in front of the sample to select circular polarization channels. Circular polarization is used to suppress single-scattering contributions in the coherent backscattering cone, which would otherwise result in a reduced enhancement factor.\cite{muskensOE08} The spectral correlation function was obtained at each angle by averaging the correlation functions from 50 spectra taken at different sample positions. The correlation function was averaged over a spectral range from 550-850~nm, yielding an average over several thousands of individual speckles. The configuration could be switched between a transmission geometry and a backscattering geometry as illustrated in Fig.~\ref{fig:setup}. The backscattering setup had an angular resolution of 1.2~mrad. For large-angle backscattering measurements, the beam-splitter was replaced by a small reflecting prism following Ref.~\onlinecite{muskensOE08}. For the transmission experiments, the beam waist at the sample position was varied by using a set of lenses of focal lengths ranging between 5~cm and 60~cm. The resulting beam waist at the sample position was determined using a knife edge method, yielding values of $\rho_0$ between $26.4 \pm 0.7$~$\mu$m and $328 \pm 16$~$\mu$m.

For the transmission measurements in Sec.~\ref{sec:expT} we used two thin slabs of TiO$_2$ powder with thicknesses of 3.4~$\mu$m and 6.3~$\mu$m. The mean free path of the sample ranges between 0.55-0.9~$\mu$m in the visible part of the spectrum.\cite{muskensOE08} The thin TiO$_2$ slabs have been investigated before using time-resolved transmission, and were found to be within the limits of classical diffusion description.\cite{Vellekoop} In the reflection experiments of Sec.~\ref{sec:expR} we study a series of random scattering media with different inverse photonic strengths $k_{e} \ell_B$, covering nearly an order of magnitude from $k_{e} \ell_B = 3.6 \pm
0.8$ to $k_{e} \ell_B = 26 \pm 3$. The materials under study include samples of etched porous GaP which have been under investigation earlier for the possible effects of localization.\cite{JohnsonPRB03,SchuurmansPRL99} These samples have been shown to be of sufficient thickness and with a low residual absorption for sensitive measurements of long light paths. In addition, we also include in our studies two layers of GaP nanowires, which have been shown to be among the most strongly scattering nanomaterials available today.\cite{muskensNanoLett09}

\section{Experiments}

\subsection{Frequency correlations in transmission}
\label{sec:expT}

\begin{figure}[t]
\includegraphics[width=8.5cm]{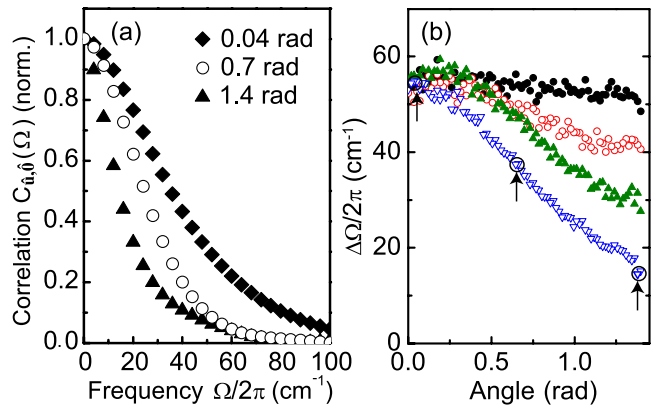}
\caption{\label{fig:c1angle} (color online) (a) Experimental frequency correlations after transmission through a 3.4-$\mu$m slab of TiO$_2$ powder, for a few typical forward angles. (b) Experimental correlation width $\Delta \Omega/2\pi$ against transmission angle for a beam waist $\rho_0$ of $26.4\pm0.7$~$\mu$m (dots, black), $55\pm
2$~$\mu$m (circles, red), $120\pm 4$~$\mu$m (upright triangles,
green), $328 \pm 16$~$\mu$m (open triangles, blue). The three black open circles with arrows correspond to the experiments displayed in (a).}
\end{figure}

\begin{figure}[t]
\includegraphics[width=7.0cm]{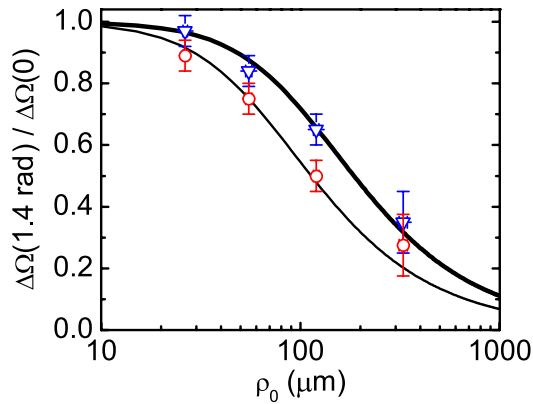}
\caption{\label{fig:widthvsrho} (color online) (a) Experimental
values of the spectral correlation width $\Delta \Omega$ at 1.4~rad, normalized to the width at 0~rad, against beam waist $\rho_0$, for TiO$_2$ slabs of 3.4~$\mu$m (circles, red) and 6.3~$\mu$m (triangles, blue) thickness. (lines) Calculated
spectral narrowing of the $C^{(1)}_T$ correlation of Eq.~(\ref{eq:c1t}) for transmission.}
\end{figure}

We investigated experimentally the angle-dependence of the frequency correlation in transmission for various values of the beam waist $\rho_0$. Figure~\ref{fig:c1angle}(a) shows typical spectral correlation functions for  $\rho_0=328 \pm 16$~$\mu$m at transmission angles of 0.04, 0.7, and 1.4~rad, 0.0~rad indicating the forward scattering direction. The exact forward direction was not measured as this contained contributions from the coherent beam for the optically thin samples ($L/\ell_B\simeq 5$ for the 3.4~$\mu$m thick slab). The correlation functions were normalized to the second data point, to remove any uncorrelated noise which is accumulated at zero frequency shift. A pronounced narrowing of the spectral correlation is observed for increasing transmission angles. This narrowing is further characterized in Fig.~\ref{fig:c1angle}(b) where the full-width-at-half-maximum $\Delta \Omega$ of the spectral correlation is plotted against transmission angle, and in Fig.~\ref{fig:widthvsrho} showing the narrowing at 1.4~rad for the two different slabs. Clearly, the narrowing depends on beam waist $\rho_0$, in agreement with our theoretical calculations using Eq.~(\ref{eq:c1t}), as illustrated by the lines in Fig.~\ref{fig:widthvsrho}. For the calculations we used values of $L$, $\ell$, and $D$ obtained from other experiments and which produced the correct width of the frequency correlation function at zero angle. Therefore, the curves in Fig.~\ref{fig:widthvsrho} do not contain any adjustable parameters. The increase of the slab thickness from 3.4~$\mu$m to 6.3~$\mu$m results in a longer dwell time and thus in a reduction of the correlation width. This results in a correspondingly less strong angle-dependence, following Fig.~\ref{fig:narrowing}. We point out that the agreement with theory is limited to beam sizes that are smaller than the diameter of our detector, which in our experiments is set by the 500~$\mu$m diameter fiber. We conclude that the model including angle-dependent propagation outside the medium, as given by Eq.~(\ref{eq:c1t}), provides a good quantitative description of our experimental results, including the scaling of the effect with $\rho_0$.

\subsection{Frequency correlations in coherent backscattering}
\label{sec:expR}
\begin{figure}[t]
\includegraphics[width=8.5cm]{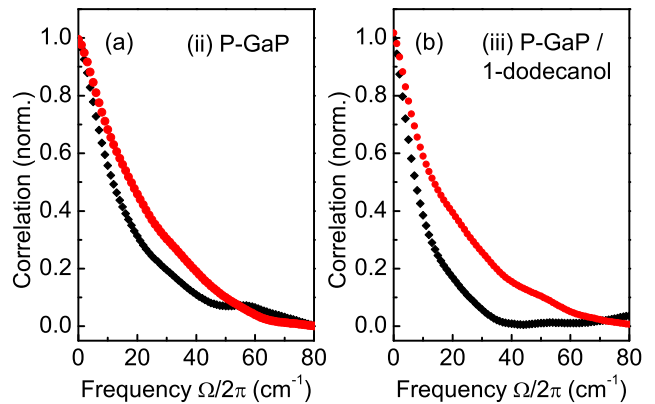}
\caption{\label{fig:c1traces} (color online) Experimental correlation functions in reflection (normalized) for porous GaP samples before (a) and after (b) infiltration with 1-dodecanol. Symbols indicate correlation in backscattering (diamonds, black at $\theta=0$~rad) and at the maximum broadening (dots, red at respectively $\theta=-0.06$~rad and $\theta=0.024$~rad). (ii), (iii) correspond to Fig.~\ref{fig:c1exp}.}
\end{figure}

In the following, we examine the single-channel frequency correlation around the coherent backscattering direction. Compared to the transmission experiments, the reflection measurements require slabs of sufficient thickness and scattering strength to resolve the coherent backscattering effect. We have chosen to study a series of slabs of porous GaP with different scattering strengths, as these have been shown to cover the transition from diffuse scattering to the strong scattering regime near localization.\cite{SchuurmansPRL99, JohnsonPRB03} In addition, two strongly scattering porous GaP slabs were measured before and after infiltration with 1-dodecanol. Infiltration with 1-dodecanol lowers the photonic strength by a factor three; this method allows unambiguous separation of the effects of scattering strength from other, sample dependent contributions.\cite{SchuurmansPRL99} Figure~\ref{fig:c1traces} shows the experimental correlation functions of a porous GaP sample before (a) and after (b) infiltration, and for angles corresponding to exact backscattering ($\theta=0$~rad) and angles where the largest change in spectral width occurs. The frequency correlation functions in reflection show a triangular peak which is slightly rounded by our spectral resolution of around 3~cm$^{-1}$. Upon infiltration with 1-dodecanol, the frequency correlation narrows indicating an increase of the photon dwell times resulting from the increase in mean free path.

\begin{figure}[t]
\includegraphics[width=8.3cm]{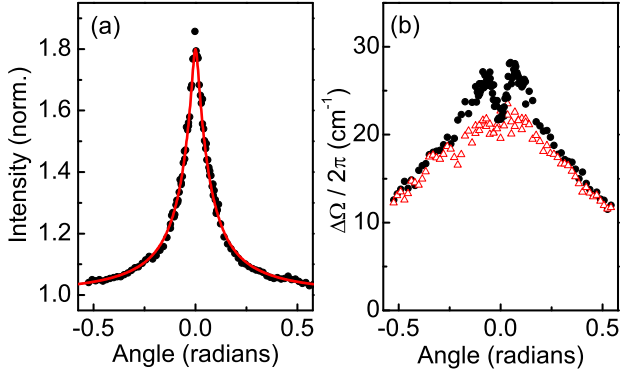}
\caption{\label{fig:c1expcbsf5} (color online) (a) Experimental intensity CBS-cone measured for
a porous GaP slab, with (line) fit for $k_{e} \ell_B = 4.7 \pm 0.6$. (b) Frequency
correlation $1/e$ half-width $\Delta \Omega/2\pi$
against backscattering angle for circularly polarized conserving
(dots, black) and linearly polarized nonconserving (red, open
triangles) channels.}
\end{figure}

In the same experimental configuration as used for the frequency correlations, conventional intensity CBS cones were obtained by collecting the average intensity. Figure~\ref{fig:c1expcbsf5}(a,b) shows a typical combination of  intensity CBS cone (a) and the $1/e$ frequency half-width of the spectral correlation function (b) as a function of backscattering angle. The black dots represent measurements taken in the helicity conserving channel, while the open triangles (red) in Fig.~\ref{fig:c1expcbsf5}(b) are measurements in the polarization nonconserving channel. The narrowing of the spectral correlation at large angles can be attributed to the dephasing contribution of Eq.~(\ref{eq:Ifarangle2}). As the path length distribution in reflection is governed by short light paths of the order of the mean free path $\ell_B$, this angle-dependent narrowing stays prominent even when $L$ increases to infinity, i.e. for semi-infinite slabs.

\begin{figure}[t]
\includegraphics[width=8.7cm]{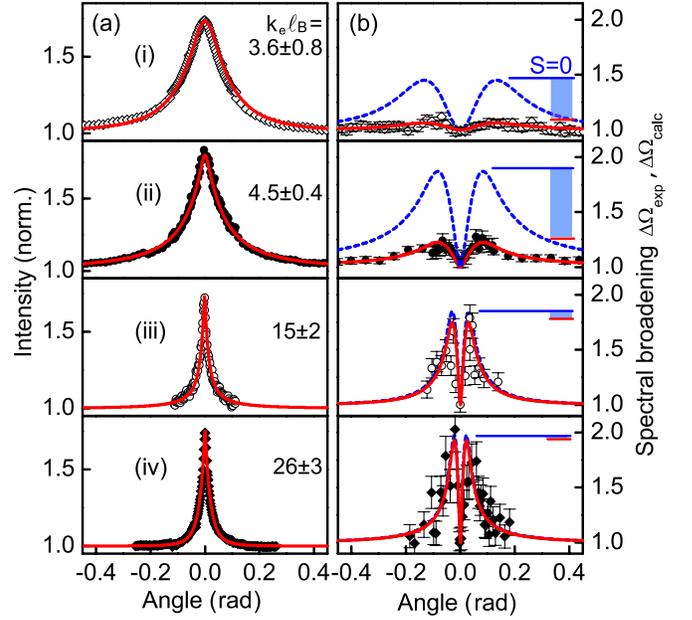}
\caption{\label{fig:c1exp} (color online) Left (a) Conventional intensity-CBS cones for different random
scattering media, with from top to bottom: compressed nanowires,
photoanodically etched porous GaP in air and in dodecanol, and etched
porous GaP, with values $k_{e}\ell_B$ ranging from $3.6 \pm 0.8$ to $26 \pm 3$. (Lines, red) Theoretical fits of conventional CBS \cite{Mark88}. Right (b) Experimentally determined
increase in the width of the spectral correlation function $\Delta \Omega_{\rm exp}$. (Lines) Theoretical calculations of
using Eq.~(\ref{eq:c1bistatic}) for known material parameters ($\Delta \Omega_{\rm calc}$, dashed lines, blue), and including
deviation parameter $S$ characterizing the long range correlations ($\Delta \Omega_{\rm exp}$, lines, red). Vertical bar denotes deviation from theoretical maximum ($S=0$, horizontal line, blue).}
\end{figure}

An additional feature is observed in the helicity conserving channel which resembles the calculated response of Fig.~\ref{fig:c1cbscalc}(d). For the narrow angular range around backscattering where the effects of the CBS cone are most pronounced, the additional narrowing due to dephasing is relatively small. We therefore analyze the broadening associated with the coherent backscattering effect by defining an experimental spectral broadening parameter $\Delta \Omega_{\exp}(\theta)$ as the ratio of the helicity conserving and nonconserving widths $\Delta \Omega$ at every angle $\theta$. Results are shown in Fig.~\ref{fig:c1exp}(a,b) for 4 different samples with scattering
strength ranging from $k_{e} \ell_B = 26 \pm 3$ to $k_{e} \ell_B = 3.6 \pm
0.8$. The wavevector $k_{e}$ includes the effective
refractive index of the scattering medium, $n_{e}$, ranging between 1.4 and
2.0 for the materials under study. Backscattering corresponds to $\theta=0.0$~rad. Conventional CBS cone (i) corresponds to the most strongly scattering nanowire material of
Ref.~\onlinecite{muskensNanoLett09}. CBS cones (ii) and (iii)
in Fig.~\ref{fig:c1exp}(a) were taken from the same porous GaP sample
before and after infiltration with 1-dodecanol, consistent with earlier measurements on the same sample.\cite{SchuurmansPRL99} Cone (iv) corresponds to a less strongly scattering porous GaP layer. The red lines in
Fig.~\ref{fig:c1exp}(a) show fits using a finite-slab
model\cite{Mark88} including an internal reflection correction,
yielding values of $k_{e} \ell_B$ as indicated in the figure. All conventional CBS cone
shapes can be described by the finite sample thickness or limited angular resolution, the estimated absorption lengths exceed the sample size.\cite{SchuurmansPRL99}

The characteristic angle-dependence of
$\Delta \Omega_{\rm exp}$ is found in Fig.~\ref{fig:c1exp}(b) for all samples under study, however with different magnitudes of the broadening effect. The angle-dependent broadening can be reproduced theoretically using Eq.~(\ref{eq:c1bistatic}) using the known sample parameters (dashed lines, blue). This model gives good quantitative agreement for the least strongly scattering material with $k_e \ell_B=26$. For the more strongly scattering
samples with $k_{e}\ell_B<10$, the magnitude of the experimental broadening is significantly
smaller than would be expected on the basis of the $C^{(1)}$
theory. In our model analysis, we take into account the effect of known finite slab thicknesses, which cuts off of light paths above a critical length in the propagator of Eq.~(\ref{eq:Hslab}). This correction is especially important for the nanowire material (i) for which $L/\ell \simeq 25$.

\begin{figure}[t]
\includegraphics[width=7.0cm]{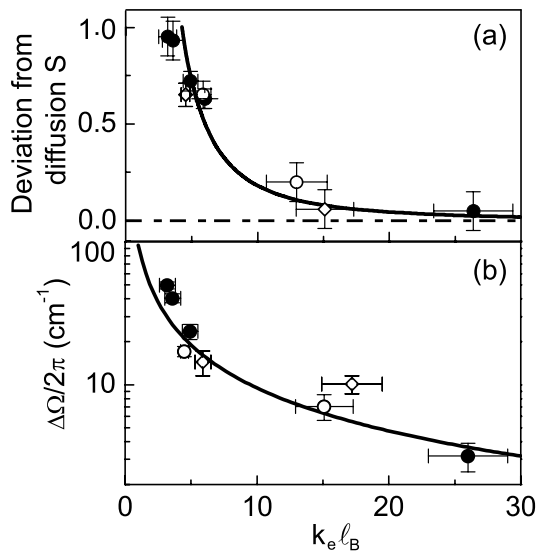}
\caption{\label{fig:c1expkl} (a) Experimentally observed deviation from the simple $C^{(1)}$ CBS model, $S$, against $k_{e} \ell_B$. (Open symbols) Pairs of
scattering samples before and after infiltration with dodecanol.
Lines represent $C^{(1)}$ value of $S=0$ (dash-dotted line) and Eq.~(\ref{eq:C2Stephen}),
$S=\frac{27}{2}(k_{e}\ell_B)^{-2}$ (solid line). (b) Spectral $1/e$ half width $\Delta \Omega$ of the $C^{(1)}$ correlation function at exact backscattering $\theta=0$ against $k_{e}\ell_B$ for the
different materials under study. Line denotes fit using the
classical diffusion theory with $\Delta \Omega \simeq D/\ell_B^2 \propto 1/k_{e}\ell_B$. }
\end{figure}

The red lines in Fig.~\ref{fig:c1exp}(b) correspond to the $C^{(1)}$ theory with an amplitude fitted to match the experimental data. We find that the theory can be scaled to our experimental data to good agreement using only a single, angle-independent scale factor. Therefore, in order to quantify disagreement with the diffusion model, we introduce an angle-independent parameter $S$ describing the percentage of deviation from diffusion, which is defined as the ratio of experimental to calculated broadening curves, offset by the baseline of unity, i.e.
\begin{equation}
S\equiv \frac{\Delta \Omega_{\rm
calc}^{\rm max}-\Delta \Omega_{\rm
exp}^{\rm max}}{\Delta \Omega_{\rm
calc}^{\rm max}-1} \, .
\end{equation}
The finite values of the slab thickness results in reduced amplitudes of the calculated curves, as can be observed in Fig.~\ref{fig:c1exp}(b).The vertical bars (blue) in Fig.~\ref{fig:c1exp}(b) indicate the deviation of the experimental results from the $C^{(1)}$ theory corrected for slab thickness ($S=0$, blue lines). The resulting deviation from classical diffusion $S$ is presented for all studied
samples in Fig.~\ref{fig:c1expkl}(a) as a function of the inverse
photonic strength $k_{e} \ell_B$. The experimental data show a strong
increase of $S$ toward $k_{e} \ell_B \simeq 3.6$, which is unaccounted
for by finite sample thickness and absorption, as can be concluded from the CBS-intensity cones of Fig.~\ref{fig:c1exp}(a). Infiltration with
dodecanol significantly reduces the deviation from classical diffusion, as indicated by the open symbols in Fig.~\ref{fig:c1expkl}. Each type (hollow circles, diamonds) represents one particular sample before and after infiltration. The pronounced effect of infiltration agrees with our expectation for an effect depending on the scattering strength, rather than on other sample-dependent properties. The frequency width $\Delta \Omega$ of the correlation in
exact backscattering, shown in Fig.~\ref{fig:c1expkl}(b), follows a $1/k_{e}\ell_B$ dependence which can be interpreted from classical diffusion in the following way. For classical diffusion, the frequency width $\Delta \Omega$ is proportional to $D/\ell_B^2$, which in combination with $D=v_E \ell_B/3$ yields the observed inverse linear dependence on $\ell_B$. A renormalization of the diffusion
constant\cite{SkipetrovPRL06,Cheung04} would lead to a narrowing
of the frequency correlation, which is not observed in our
experiment.

\section{Discussion}

We interpret the general trend in Fig.~\ref{fig:c1expkl}(a)
starting from the diffuse scattering regime. In the CBS cone, the
wings of the cone correspond to very short transverse distances
between the point of entry and the point of exit of light in the
medium, of the order of the mean free path $\ell_B$.\cite{Mark88}
The frequency broadening $\Delta \Omega_{\rm exp}$ in the wings is
sensitive to the phase delay of these short paths relative to that
of the total distribution. In the $C^{(1)}$ approximation, the
maximum frequency broadening $\Delta \Omega_{\rm calc}$ is
independent of $k_{e}\ell_B$, as indicated by the dash-dotted line
in Fig.~\ref{fig:c1expkl}(a). The deviation of our experimental data from this line can be caused by two possible types of effects, which either affect the dwell time, or which add new contributions to the spectral correlation. To the first type belongs the cutoff of long light paths and the associated distribution of decay times near localization.\cite{SchuurmansPRL99,SkipetrovPRL06} The second type includes higher order contributions to the correlation function known as $C^{(2)}$.\cite{Feng94} Several authors have considered this contribution in a reflection geometry.\cite{Genack90,Stephen87,Wang,Berkovits90,Berkovits90b,Rogozkin1995} The angular $C^{(2)}$ correlation is inversely proportional to the illumination area, and is expected to be negligibly small for the geometry of our experiment. In contrast, the near-field correlation in reflection is a local effect on a length scale of order $\ell$, which does not vanish for plane wave illumination. This long-range near-field correlation has been calculated by Stephen and Czwilich as\cite{Stephen87}
\begin{equation}\label{eq:C2Stephen}
C^{(2)}(R) \simeq
\frac{27}{2k_{e}^2\ell_B^2}\left(\frac{\ell_B}{R}\right)^3.
\end{equation}
Equation~(\ref{eq:C2Stephen}) shows that, for short the distances $R\simeq \ell_B$, contributing to the wings of the CBS cone, the near-field intensities are much stronger correlated than for distances $R>>\ell_B$ contributing to the center of the cone.

\begin{figure}[t]
\includegraphics[width=7.5cm]{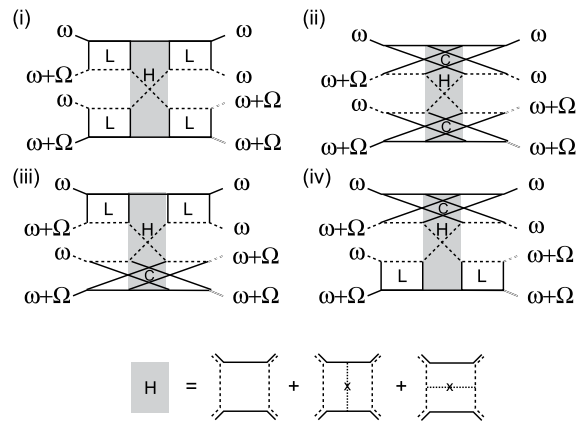}
\caption{\label{fig:hikami} Higher-order diagrams in the frequency correlation in reflection, including time-reversal symmetry \cite{Berkovits90b}. L and C are ladder and Cooperon vertices, H indicates the Hikami-box vertex consisting of three terms.\cite{AkkermansBook}}
\end{figure}

At this stage, a full theoretical framework of the $C^{(2)}$ contribution to the coherent backscattering frequency correlation has yet to be developed. Figure~\ref{fig:hikami} shows the higher-order diagrams that have to be included in such an extended model, following Ref.~\onlinecite{Berkovits90b}. The four-field correlation takes place through the Hikami-box vertex, which is not present in the conventional intensity CBS. In addition, interference with an unscattered, reduced incident intensity may be important.\cite{Rogozkin1995} In a different context, these 'shadow' terms have been identified for their role in energy conservation in the CBS-cone.\cite{Maret08} Althogether, the frequency correlation CBS may contain new information compared to the conventional intensity CBS, where $C^{(2)}$ is only a background.\cite{WiersmaPRL95} This is important as our measurements show a collapse of the CBS effect in the correlation function, while the intensity CBS cone is still intact. A crucial point is that self-interference due to the intersection of paths is associated with a finite intersection volume of order $\lambda^2 \ell_B$, with $\lambda$ the optical wavelength.\cite{AkkermansBook} This is especially relevant in the wings of the CBS cone, where the transverse distance between the path ends is only one mean free path $\ell_B$. Following the scaling of Eq.~\ref{eq:C2Stephen}, the condition $C^{(2)}(\ell_B)=1$ is reached when $k_{e}\ell_B=(27/2)^{1/2}\simeq 3.67$. The breakdown of the weak-scattering limit for the $C^{(1)}$ correlation is thus expected to occur far above the localization limit $k_{e}\ell_B \sim 1$ in the wings of the CBS cone. The line in Fig.~\ref{fig:c1expkl}(a) illustrates this scaling, where we hypothesized a deviation $S$ of zero in the diffuse limit up to full breakdown $S=1$ of the $C^{(1)}$ model due to path intersection.

\begin{figure}[t]
\includegraphics[width=8.0cm]{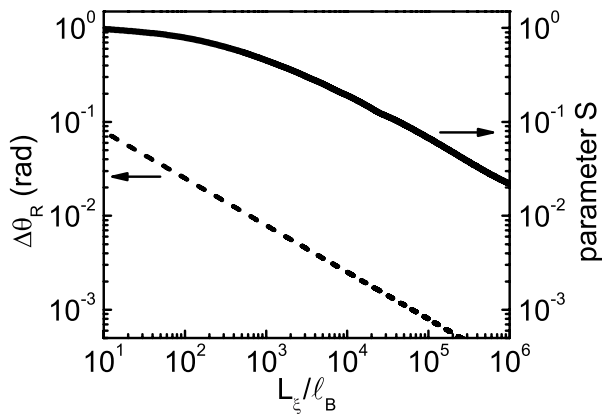}
\caption{\label{fig:comparisons} Comparison between the sensitivity of the parameter $S$ and the CBS cone rounding\cite{SchuurmansPRL99} $\Delta \theta_R$ to the suppression of long light paths, represented by the cutoff length $L_\xi$.}
\end{figure}

Apart from possible higher-order contributions in the spectral correlation, we consider the sensitivity of the CBS-induced broadening of the frequency correlation to changes in the path length distribution. In particular, we compare the new technique to measurements of the CBS-cone rounding in intensity measurements.\cite{SchuurmansPRL99} Figure~\ref{fig:comparisons} shows both the cone rounding angle $\Delta \theta_R$ and the suppression of the frequency broadening $S$, as a function of the path length cutoff defined by a (total) length $L_\xi$. For simplicity we choose an exponential cutoff such as is the case for absorption in Eq.~(\ref{eq:Hslab}), however alternative choices of the path length scaling are available to describe other effects such as localization.\cite{SkipetrovPRL06} Experiments on the same porous GaP samples have identified an anomalous cone rounding $\Delta \theta$ of around 4~mrad.\cite{SchuurmansPRL99} In Fig.~\ref{fig:comparisons} this value corresponds to a value $L_\xi/\ell$ of $3.7\times 10^3$, or a cutoff length $L_\xi$ of around 1~mm for the porous GaP sample. In comparison, for the same $L_\xi/\ell$ the parameter $S$ gives a value of 0.28. Experimentally we have observed, for the same sample as used in Ref.~\cite{SchuurmansPRL99}, an $S$-value of $0.65\pm 0.04$. Recent experiments have reported sizeable corrections to diffuse transport for values of $k\ell_B$ larger than 1 (respectively $k\ell_B\simeq 2.5$ and $k\ell_B\simeq 1.8$ in Refs.~\onlinecite{StoerzerPRL06} and ~\onlinecite{Page08}). Self-consistent theories of time-dependent diffusion\cite{SkipetrovPRL06, Cheung04} predict a reduction of the diffusion constant in this regime, which should lead to a narrowing of the spectral correlation. This is not observed experimentally in Fig.~\ref{fig:c1expkl}(b). However, we point out that both the horizontal axis depending on $\ell_B$ and the vertical axis depending on $D$ should be renormalized in this regime, making it hard to observe these renormalizations experimentally as there is no absolute reference against which these can be calibrated.

\section{Conclusions}

In conclusion, we have studied both theoretically and experimentally the angle-dependence of frequency correlations in random photonic media both in transmission and in reflection. A narrowing of the spectral correlation was found which depends on both the scattering angle and the incident beam waist, which could be identified as an additional dephasing of the multiple scattered light outside the medium. This contribution has not been reported in other work on frequency correlations.\cite{Feng94,Genack87,Genack90,DeBoer92,Albada90} We note that earlier experimental studies dealt with much thicker slabs for which the width of the frequency correlation was too small to observe the effect according to Fig.~\ref{fig:narrowing}. However, the angle-dependent narrowing is sizeable for frequency widths above 10~cm$^{-1}$, which corresponds to dwell times below 1~ps. Thus the effect is relevant for many nanophotonic materials and thin films in which light trapping plays a role, and particularly in reflection geometries where the path length distibution is governed by short paths on the scale of the mean free path $\ell_B$, even for very thick samples.

In backscattering, we have found a contribution to the frequency correlation which can be attributed to the angle-dependent path length distribution in coherent backscattering. Our diffusion model is confirmed by experiments for scattering materials with inverse photonic strength $k_e \ell_B\simeq 26$, however the magnitude of the broadening effect appears reduced for samples with larger scattering strength. We emphasize that a similar strong dependence is not found for the intensity CBS cones obtained in the same experiment. Thus, the frequency correlation in backscattering may form a new instrument for accessing the breakdown of diffusion theory in the limit of very strong multiple scattering. Earlier work on similar (and some of the same) strongly scattering materials revealed no or only weak corrections to classical diffusion.\cite{JohnsonPRB03} It has been pointed out that time-resolved transmission is not expected
to yield large modifications for the typical samples under study \cite{SkipetrovPRL06}. The angle-dependent frequency correlation provides a means for accessing the path length distribution in reflection on much shorter time scales than can be accessed using ultrafast laser pulses. The frequency-resolved technique thus yields information complimentary to picosecond time-resolved experiments which are usually more involved and not easily extended to studying angle-dependent phenomena.\cite{Vreeker89,Toninelli08,StoerzerPRL06,JohnsonPRB03} The frequency correlations in coherent backscattering have the potential to be an even more precise probe of changes in the path length distribution than cone rounding measurements using the intensity CBS, making it an important tool for studying the strong scattering regime near localization. Our results call for new theoretical models for frequency correlations in combination with coherent backscattering in the strong scattering limit.

\section{Acknowledgements}
This work is part of the research program
of the "Stichting voor Fundamenteel Onderzoek der Materie (FOM)",
which is financially supported by the "Nederlandse Organisatie
voor Wetenschappelijk Onderzoek (NWO)".

\newpage

%%%%%%%%%%%%%%%%%%%%%%%%%%%%%%%%%%%%%%%%%%%%%%%%%%%%%%%%%%%%%%%

\end{document}